# GEM: A GEneral Memristive Transistor Model

Shengbo Wang[1], Jingfang Pei[2], Cong Li[1], Xuemeng Li[1], Li Tao[3], Arokia Nathan[4,5], Guohua Hu[2] and Shuo Gao[1], *Senior Member, IEEE*
[1]*School of Instrumentation and Optoelectronic Engineering, Beihang University, Beijing, China*
[2]*Department of Electronic Engineering, The Chinese University of Hong Kong, Shatin, N. T., Hong Kong SAR, 999077 China*
[3]*School of Physics, Beijing Institute of Technology, Beijing, China*
[4]*Darwin College, University of Cambridge, Cambridge, UK*
[5]*School of Information Science and Engineering, Shandong University, Qingdao, China*
shuo_gao@buaa.edu.cn

***Abstract*—Neuromorphic devices, with their distinct advantages in energy efficiency and parallel processing, are pivotal in advancing artificial intelligence applications. Among these devices, memristive transistors have attracted significant attention due to their superior stability and operation flexibility compared to two-terminal memristors. However, the lack of a robust model that accurately captures their complex electrical behavior has hindered further exploration of their potential. In this work, we introduce the GEneral Memristive transistor (GEM) model to address this challenge. The GEM model incorporates time-dependent differential equation, a voltage-controlled moving window function, and a nonlinear current output function, enabling precise representation of both switching and output characteristics in memristive transistors. Compared to previous models, the GEM model demonstrates a 300% improvement in modeling the switching behavior, while effectively capturing the inherent nonlinearities and physical limits of these devices. This advancement significantly enhances the realistic simulation of memristive transistors, thereby facilitating further exploration and application development.**

## I. Introduction

Neuromorphic devices, with their inherent advantages in energy efficiency and parallel processing capabilities, have emerged as promising candidates for various applications, particularly in accelerating artificial intelligence algorithms [1]-[7]. Among these devices, memristive (or synaptic) transistors have garnered significant attention due to their superior stability and operation flexibility compared to traditional two-terminal memristors [8]-[12].

Despite these advantages, the lack of a model that accurately captures the behavior of memristive transistors has hindered effective simulation and exploration of their applications. Although existing models like VTEAM have successfully modeled switching behaviors in two-terminal memristors, they fall short in capturing the essential electrical characteristics of memristive transistors, including the range of state changes under varying modulation voltages and output characteristics across multiple operating regions [13]-[15].

To bridge this gap, we propose the GEneral Memristive Transistor (GEM) model, which characterizes state switching dynamics through a time-dependent differential equation. Additionally, GEM integrates a voltage-controlled moving window to restrict its state range and a nonlinear current output function to reflect the intrinsic output properties of transistors. This approach allows GEM to simulate memristive transistors with different structures and switching mechanisms at a behavioral level. Our experiments validate the generality of GEM by simulating practical data from various published memristive transistor types. Compared to previous models, GEM demonstrates higher accuracy in modeling switching characteristics (300% improvement) and can simultaneously capture the physical limitations and nonlinear output characteristics of memristive transistors. Thus, the GEM model paves the way for more reliable memristive transistor simulations and application exploration. The GEM model in SPICE simulator, along with the fitting code and materials, can be found at https://github.com/RTCartist/GEM-A-GEneral-Memristive-transistor-model.

The rest of this brief is organized as follows: Section II discusses current models for memristive devices and their limitations in replicating memristive transistor behavior. Section III details the GEM model, followed by a presentation of experimental results and comparisons with previously proposed models in Section IV. Finally, a summary of the brief is provided in Section V.

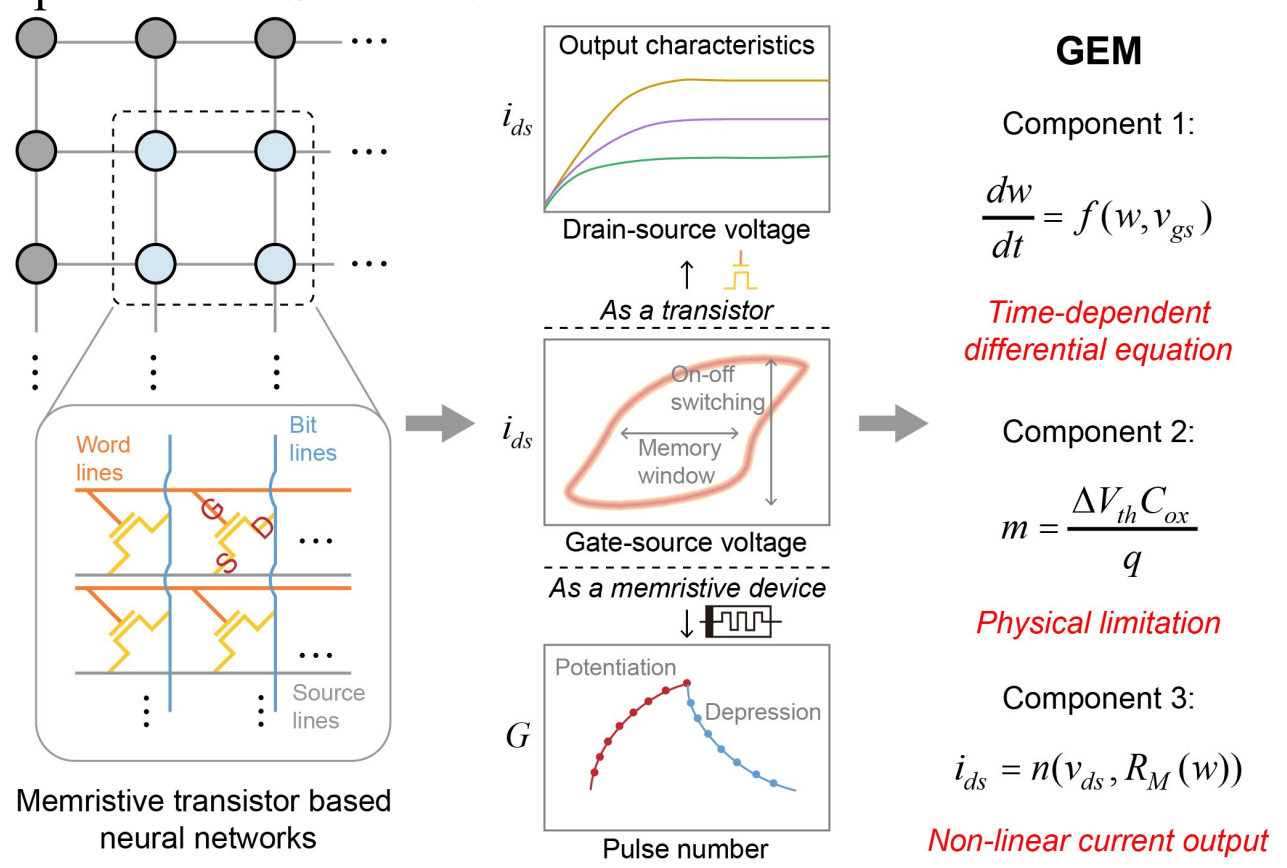

**Fig. 1.** Electrical characteristics of memristive transistors, along with the GEM model designed to simulate their behavior. Here, $i_{ds}$ denotes the drain-source current, and $G$ represents the channel conductance.

## II. Current Models and Their Limitations

### *A. Models for Memristive Devices*

Several models have been developed to describe the switching dynamics of two-terminal memristors, a widely studied memristive device. Among these, the linear ion drift model assumes a simple linear relationship between ionic drift and the externally applied electric field [16]. However, due to its simplicity, this model is insufficient for capturing the complex, non-linear switching behavior observed in physical memristors. The non-linear ion drift model improves upon this by incorporating a non-linear drift term, making it more accurate than the linear version. However, it still fails to account for the threshold voltage observed in many memristors. In this context, the TEAM and VTEAM models proposed by Kvatinsky et al. have become some of the most widely used models due to their flexibility and accuracy [17], [13], [14]. These models incorporate a threshold setting and a flexible function to describe the electrical characteristics of memristors at a behavioral level, ensuring generality without becoming computationally inefficient.

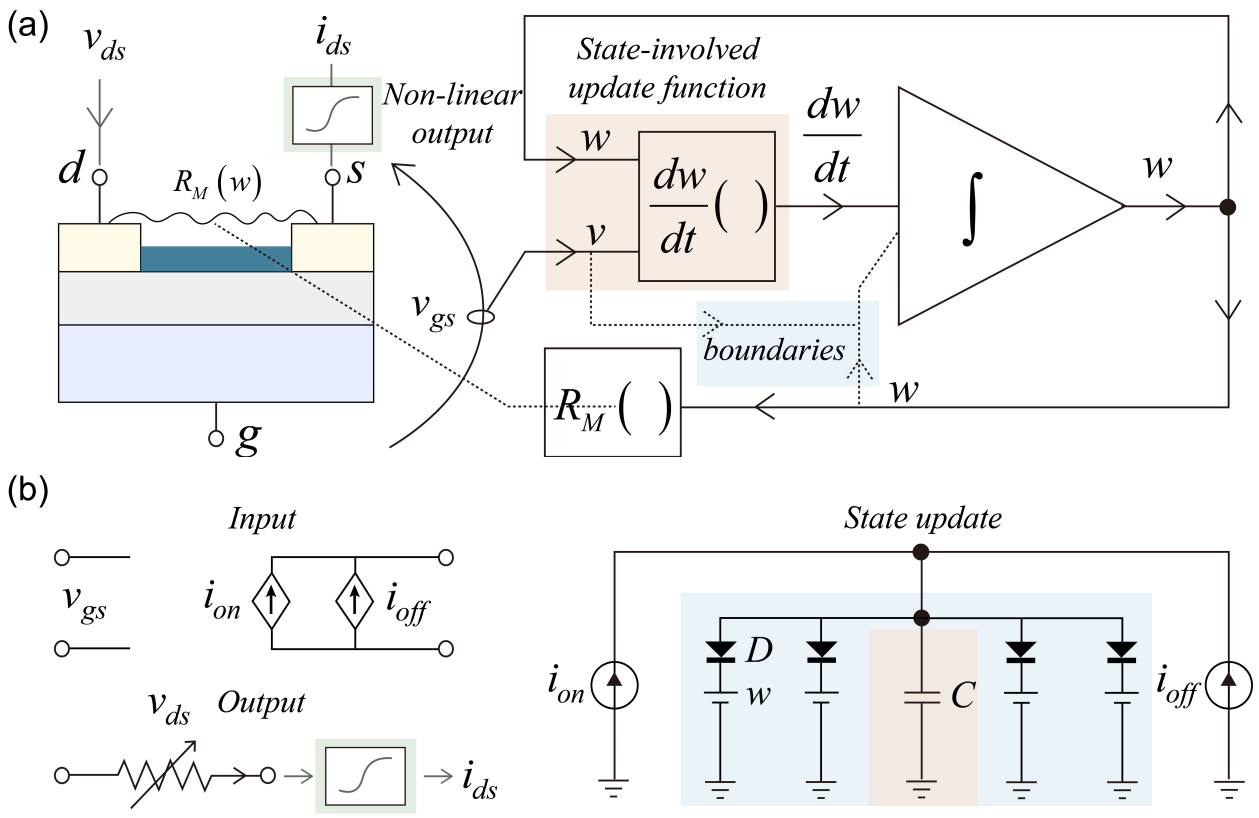

**Fig. 2.** (a) Block diagram of the GEM model, consisting three key features: a state-involved update function, voltage-controlled moving window function and a non-linear current output function. (b) SPICE implementation.

## B. Requirements for Modeling Memristive Transistors

When modeling three-terminal memristive transistors, two key aspects must be considered: state-switching behavior (as memristive devices) and output characteristics (as transistors). In terms of state-switching behavior, memristive transistors—whether based on phase change, charge trapping, or floating gate mechanisms—exhibit equivalent conductance changes under external modulation voltage pulses, typically applied between the gate and source terminals. This switching behavior depends not only on the applied electric field but also on the current state of the device. For instance, in charge trapping mechanisms, the concentration of charge carriers significantly affects the tunneling process, which in turn influences subsequent state switching behavior [9], [10]. However, existing models like VTEAM may not fully capture this dynamic behavior, resulting in reduced accuracy when modeling switching characteristics. Additionally, the state range of some memristive transistors during switching is related to the amplitude of the applied gate-source voltage [10], a critical factor that current memristive device models fail to address. From a transistor perspective, memristive transistors, like conventional transistors, exhibit distinct operating regions (e.g., linear and saturation) depending on the drain-source voltage. Therefore, an accurate model must reflect the difference in output characteristics across these regions to capture the intrinsic transistor behavior.

# III. GEM

To capture the switching behavior of memristive transistors at a behavioral level and ensure generality, the state switching behavior is modeled using an internal state variable, $w$, governed by a time-dependent differential equation. As shown in Fig. 2, this state variable is modulated by the voltage applied across the gate and source electrodes ($v_{gs}$). This updated state is then reflected in the current response $i_{ds}$, measured between the drain and source terminals under the applied drain-source voltage ($v_{ds}$). Specifically, the time-dependent differential equation is expressed as follows:

$$\frac{dw}{dt}=\begin{cases} k_{off}\left(\frac{v_{gs}(t)}{v_{off}}-1\right)^{\alpha_{off}}\left(1-\frac{w(t)}{w}s_{off}\right)^{\beta_{off}} f_{off}(w), v_{gs}<v_{off}<0 \\ 0, v_{off}<v_{gs}<v_{on} \\ k_{on}\left(\frac{v_{gs}(t)}{v_{on}}-1\right)^{\alpha_{on}}\left(1-\frac{w(t)}{w}s_{on}\right)^{\beta_{on}} f_{on}(w), v_{gs}>v_{on}>0 \end{cases} \quad (1)$$

where $t$ is time, $k_{off}$, $k_{on}$, $\alpha_{off}$, $\alpha_{on}$ $s_{off}$, $s_{on}$, $\beta_{off}$ and $\beta_{on}$ are constants, $v_{off}$, $v_{on}$ are threshold voltages, and $w(t)$ is current state. Notably, the threshold voltage here refers to the minimum gate-source voltage $v_{gs}$ required to induce a significant change in the device's memristive state. This should not be confused with the threshold voltage of conventional transistors, such as MOSFETs, where the threshold voltage marks the transition from the OFF state (non-conductive) to the ON state (conductive). The parameters $\alpha_{off}$ and $\alpha_{on}$ describe the voltage sensitivity of the memristive transistor, while the terms $s_{off}$, $s_{on}$, $\beta_{off}$ and $\beta_{on}$ capture the influence of current state. This state-involved design enhances the accuracy of the GEM model in capturing switching behavior. Additionally, $k_{off}$ and $k_{on}$ serve as linear constants to balance this representation. The functions $f_{off}$ and $f_{on}$ act as window functions to constrain the state variable $w$ within the bounds [$w_{off}$, $w_{on}$], where $w_{on}$ and $w_{off}$ represent the upper and lower bounds of the state variable $w$. When the state variable approaches the boundary of the window function, the change in $w$ ($dw$) will be set to zero by window functions $f_{off}$ and $f_{on}$, ensuring the state remains within defined limits.

In addition to this basic window function, the effect of state range variation under amplitude of $v_{gs}$ is also considered. For example, an increase in the absolute value of $v_{gs}$ leads to a wider state range, while a decrease results in a narrower range. This behavior can be explained by the capacitor model in the charge-trapping mechanism, where the maximum amount of trapped electrons (the state limit) is correlated with the amplitude of $v_{gs}$. Inspired by this mechanism, additional moving window functions $g_{off}(w,v_{gs})$ and $g_{on}(w,v_{gs})$ are introduced on the basis of Eq. 1 during the resetting and setting processes, respectively. These functions can be represented as:

$$g_{off}(w,v_{gs})=\begin{cases} 1, w>w_{on}+(w_{off}-w_{on})h_{off}\left(\frac{v_{gs}}{v_{\min}}\right) \\ 0, w\le w_{on}+(w_{off}-w_{on})h_{off}\left(\frac{v_{gs}}{v_{\min}}\right) \end{cases} \quad (2)$$

$$g_{on}(w,v_{gs})=\begin{cases} 1, w<w_{off}+(w_{on}-w_{off})h_{on}\left(\frac{v_{gs}}{v_{\max}}\right) \\ 0, w\ge w_{off}+(w_{on}-w_{off})h_{on}\left(\frac{v_{gs}}{v_{\max}}\right) \end{cases} \quad (3)$$

where $v_{\min}$ is the voltage that can modulate the state of memristive transistor to $w_{off}$, $v_{\max}$ corresponds to the state $w_{on}$, and the functions $h_{on}$ and $h_{off}$ are not inherently defined and can be optimized based on experimental data.

At the transistor level, the output characteristics of the GEM model is formulated as Eq. 4:

$$i_{ds} = n(v_{ds}, R_M(w)) = \begin{cases} \dfrac{v_{ds}}{R_M(w)}, v_{ds} - v_{gs} \le t(w) \\ \dfrac{v_{ds} m(v_{ds} - v_{gs} - t(w))}{R_M(w)}, v_{ds} - v_{gs} > t(w) \end{cases} \quad (4)$$

Here, $R_M(w)$ represents a memristive resistance controlled by the state variable $w$, while the function $n(v_{ds},R_M(w))$ is a non-linear function dependent on both $v_{ds}$ and $w$. The value $t(w)$ acts as a threshold determining whether the memristive transistor operates in the saturation region while the function $m$ specifically describes the nonlinear output characteristics. The function $R_M(w)$ is not inherently defined within the GEM model and can be determined based on experimental data to suit various memristive transistors. For example, a linear dependence on the state variable can be expressed as:

$$R_M(w) = R_{on} + \frac{R_{off} - R_{on}}{w_{off} - w_{on}}(w - w_{on}) \quad (5)$$

where $R_{on}$ and $R_{off}$ are the resistances when the state variable is $w_{on}$ and $w_{off}$, respectively. Alternatively, an exponential dependence on the state variable can be defined as:

$$R_M(w) = \frac{R_{on}}{e^{-\frac{\lambda}{w_{off} - w_{on}}(w - w_{on})}} \quad (6)$$

$$e^{\lambda} = \frac{R_{off}}{R_{on}} \quad (7)$$

Since the exponential case is inherently limited by its concave nature, we propose a logarithmic dependency with a convex function property to describe the switching behavior more comprehensively. The logarithmic dependency is defined as:

$$R_M(w) = \frac{R_{on} R_{off}}{(R_{off} - R_{on}) \ln((e-1)\dfrac{w - w_{off}}{w_{on} - w_{off}} + 1) + R_{on}} \quad (8)$$

where $e$ is the natural base. Similar to the function $R_M(w)$, the function $m$ is calibrated based on piratical data to ensure model generality.

In the SPICE implementation, as shown in Fig. 2b, the gate-source voltage controls two current sources, $i_{on}$ and $i_{off}$, which modulate the internal variable into $w_{on}$ and $w_{on}$, respectively. These currents are then integrated into a capacitor to realize the time-dependent differential equation in hardware, with the resulting voltage representing the total state change. Parallel ideal diodes and voltage sources are used as window functions to limit the range of this state change. Finally, a resistance $R_M(w)$ controlled by the state variable $w$, along with a nonlinear output function, produces the drain-source current $i_{ds}$.

## IV. Experimental Results

### A. Optimization Algorithms

To fit experimental data of memristive transistors into the GEM model, we propose a three-stage search algorithm to efficiently determine model parameters. These three stages include switching behavior modeling, state range optimization, and output curve determination. During the switching behavior modeling stage, this algorithm minimizes the Root Mean Square Error (RMSE) as defined in Eq. 9 by adjusting the parameters related to switching behavior.

$$RMSE = \sqrt{\frac{1}{N}\left(\frac{\sum_{i=1}^{N}(i_{GEM,i} - i_{real,i})^2}{\sum_{i=1}^{N} i_{real,i}^2}\right)} \quad (9)$$

where $i_{real}$ represents the known experimental current data used as a reference, and $i_{GEM}$ is the inferred current data based on the optimized GEM model and the practical $v_{gs}$ data. The index $i$ represents the $i$th data point among total $N$ data points. This optimization algorithm is adaptable to various types of experimental data by using different reference data types. For example, in typical synaptic behavior tests that focus on changes in equivalent conductance under $v_{gs}$ voltage pulses, the RMSE function can be refined as shown in Eq. 10.

$$RMSE = \sqrt{\frac{1}{N}\left(\frac{\sum_{i=1}^{N}(R_{GEM,i} - R_{real,i})^2}{\sum_{i=1}^{N} R_{real,i}^2}\right)} \quad (10)$$

where $R_{GEM}$ and $R_{real}$ represent the inferred and experimental resistance data, respectively.

After selecting the RMSE function, the first-stage fitting procedure iterates on parameters $k_{on}$, $k_{off}$, $\beta_{off}$ and $\beta_{on}$ to minimize the RMSE using a combination of global optimization via simulated annealing and local optimization via the Quasi-Newton method, as outlined in Algorithm 1. Other parameters, including $R_{on}$, $R_{off}$, $v_{on}$ and $v_{off}$ are set based on experimental observations. Notably, $R_{on}$ and $R_{off}$ are calculated from the observed maximum and minimum drain-source current $i_{ds}$ and the applied voltage $v_{ds}$. The remaining parameters, $\alpha_{on}$, $\alpha_{off}$, $s_{off}$ and $s_{on}$, are selected manually as constant parameters to avoid convergence to a local minimum when optimizing too many parameters.

Based on the optimized parameters for switching behavior, the moving window function can be determined by optimizing the relationship (functions $h_{on}$ and $h_{off}$) between the state range and the amplitudes of voltage pulses using the least squares method. For the output characteristics, memristive transistors are normally utilized when $v_{gs}$ is set to 0 V; therefore, Eq. 4 can be simplified as the following equation:

$$i_{ds} = n(v_{ds}, R_M(w)) = \begin{cases} \dfrac{v_{ds}}{R_M(w)}, v_{ds} \le t(w) \\ \dfrac{v_{ds} m(v_{ds} - t(w))}{R_M(w)}, v_{ds} > t(w) \end{cases} \quad (11)$$

Subsequently, the functions $t$ and $m$ can be determined from experimental data.

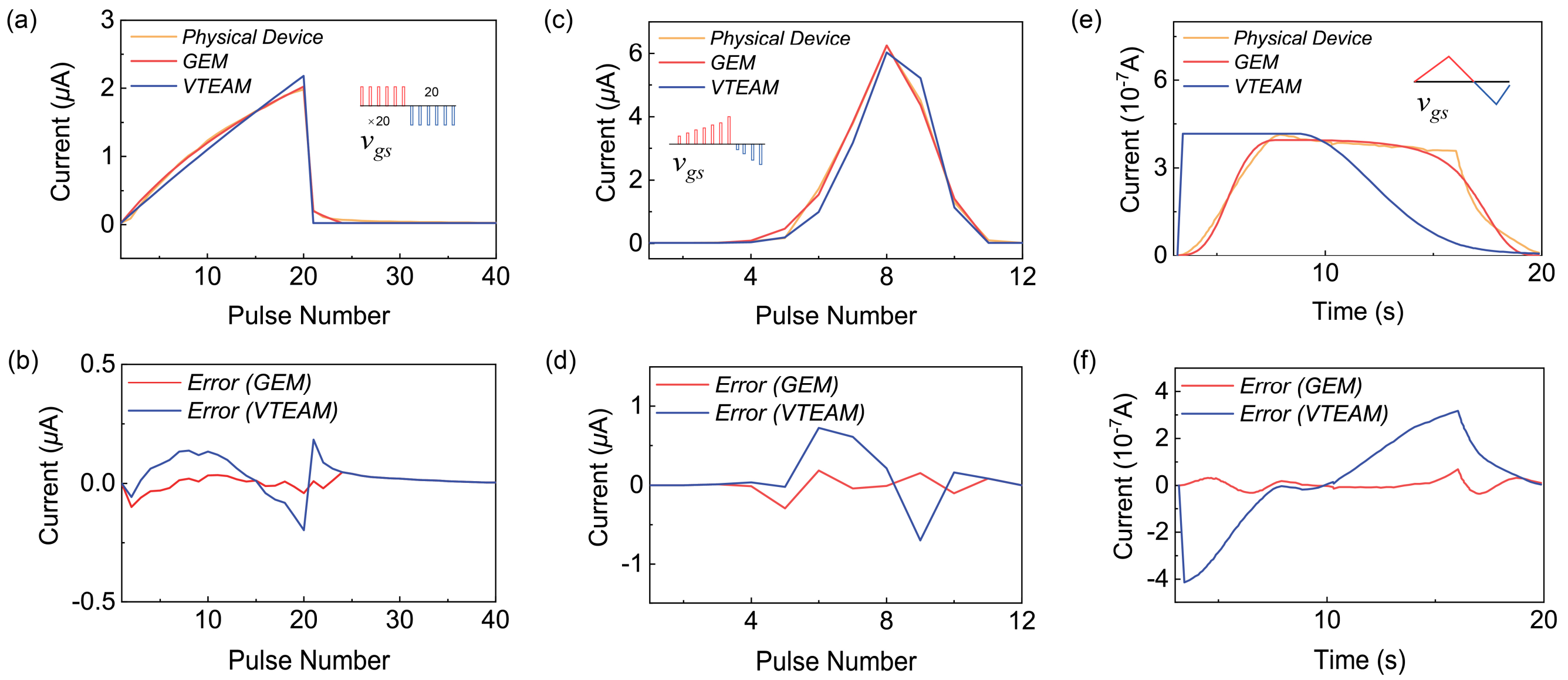

**Fig. 3.** Switching behavior optimization results. (a) Switching behavior fitting for memristive transistors described in [18]. (b) Fitting error. (c) Switching behavior fitting for memristive transistors described in [19]. (d) Fitting error. (e) Switching behavior fitting for memristive transistors described in [10]. (f) Fitting error.

**Algorithm 1** Three-Stage Optimization Method.

**Stage 1:** Switching behavior modeling

Require: Experimental data ($v_{gs}$, $v_{ds}$, $i_{ds}$), and initial parameters ($k_{on}$, $k_{off}$, $\beta_{off}$, $\beta_{on}$)

1. Initialize: $para$ = [$k_{on}$, $k_{off}$, $\beta_{off}$, $\beta_{on}$ ]
2. Error calculation: $error$ = RMSE ($i_{ds}$, $i_{model}$)
   $i_{model}$ = GEM ($v_{gs}$, $v_{ds}$, $para$)
3. Optimization: $para$ = Simulated Annealing ($error$, $para$)
   $para$ = Quasi-Newton ($error$, $para$)

return $para$ = [$k_{on}$, $k_{off}$, $\beta_{off}$, $\beta_{on}$ ] // Optimized parameters

**Stage 2:** Stage range optimization

Require: Stage range data under varying $v_{gs}$ amplitude

1. Fitting: $h_{on}$ = Least Squares ($state\ range$, $v_{gs}/v_{max}$)
   $h_{off}$ = Least Squares ($state\ range$, $v_{gs}/v_{min}$)

**Stage 3:** Stage range optimization

Require: Output characteristics data ($v_{ds}$, $i_{ds}$)

1. Fitting: [$m$, $t$] = Levenberg-Marquardt ($v_{ds}$, $i_{ds}$)

### *B. Switching Behavior Fitting Results*

In this section, three physical memristive transistors are compared with the GEM model, including a carbon nanotube synaptic transistor [18], a $WS_2$/PZT FeFET synaptic transistor [19], and a $HfS_2$/h-BN/FG-graphene synaptic transistor [10]. The optimization results are shown in Fig. 3. Compared to the current state-of-the-art general model (VTEAM) of capturing switching behavior, the GEM model demonstrates superior performance by incorporating a state-involved update function. Specifically, the RMSE for the three memristors is 0.004, 0.013, and 0.007, respectively, demonstrating an improvement of at least 300% (Table I). Additionally, the GEM model shows strong compatibility with various test data, including pulse tests and continuous signals such as triangular waves.

For the moving window function, the synaptic properties described in [18] are utilized. During testing, when varying amplitudes of positive voltage pulses are applied to the memristive transistor, the drain-source current stabilizes, with its final value proportional to the amplitude of the $v_{gs}$. Based on this data, the function $h_{on}$ is calculated as depicted in Fig. 4a. When incorporating this moving window function into the GEM model, the simulated drain-source current $i_{ds}$ is effectively constrained according to the applied voltage amplitude. As the state approaches its limit, the model achieves a mean absolute error (MAE) of 0.041 $\mu$A in current (Fig. 4b), demonstrating the ability of the GEM model to accurately implement physical limits in the simulation of memristive transistors.

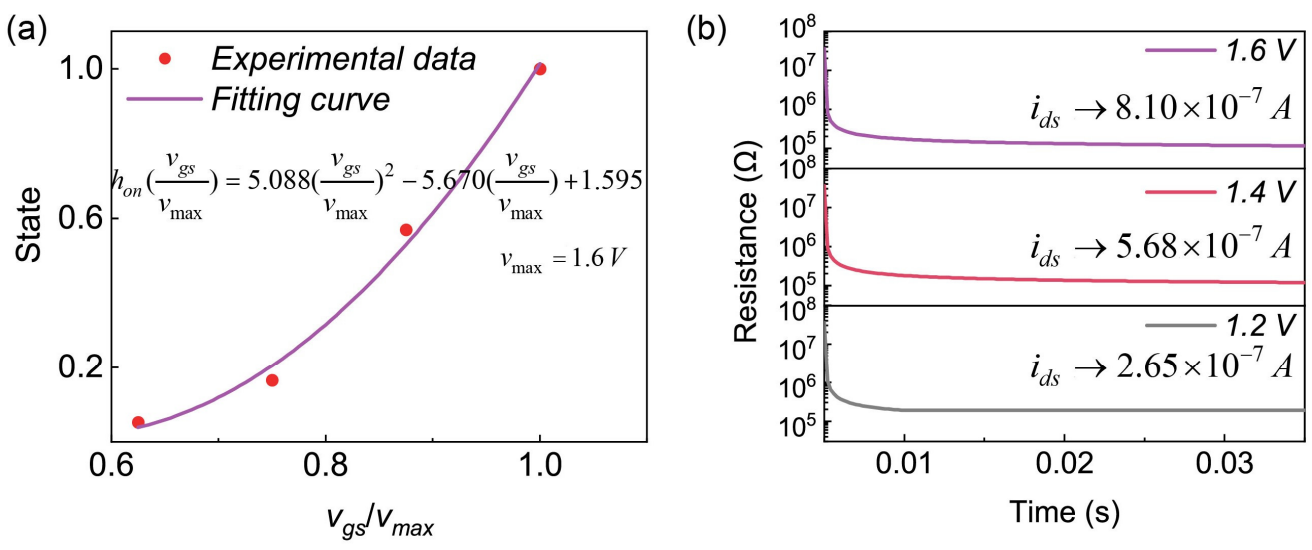

**Fig. 4.** Physical limits fitting. (a) Experimental results fitting based on second order functions. (b) Resistance change of GEM model with moving window functions under varying amplitude of voltage pulses

### *C. Output Characteristics Fitting Results*

The output characteristics curve of a three-terminal gate injection-based field-effect transistor [9] is employed to determine the non-linear output function. In this analysis, with the state of memristive transistor fixed, both $t(w)$ and $R_M(w)$ can be considered constants, and $R_M(w)$ can be derived from the slope of output curve. Notably, $t(w)$ can be flexibly obtained either through direct experimental observation or optimization algorithms. Given that the device exhibits strong Ohmic behavior below 1 V, $R_M(w)$ can be directly calculated based on the slope of output curve in this region, with $t(w)$ representing the upper bound of this region at 1 V. Subsequently, the function $m$ can be determined using the Levenberg-Marquardt algorithm (Fig. 5), and the MAE of this fitting results is 0.017 $\mu$A. When additional output curve data at different state variables are available, the functions $m$ and $t$ can be further optimized. To the best of our knowledge, GEM is the first general memristive transistor model to incorporate nonlinear output characteristics, as shown in Table I.

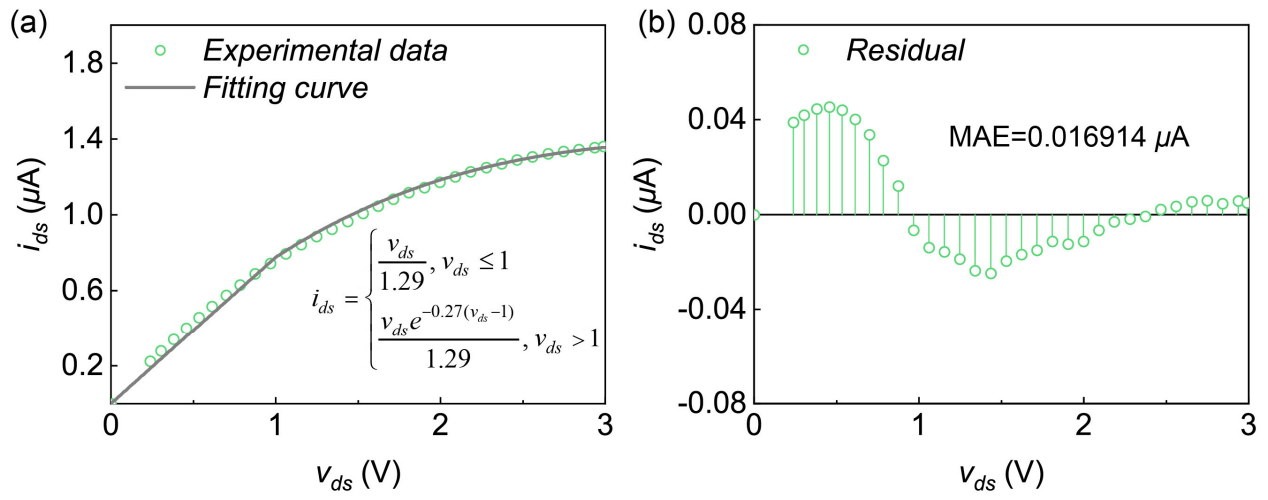

**Fig. 5.** Non-linear output function fitting. (a) Experimental results fitting. (b) Residuals of the fitting results.

TABLE I. COMPARISON WITH OTHER WORKS

| Model | **GEM** | VTEAM | Biolek |
|---|---|---|---|
| Switching capture error (RMSE) [18] | 0.004 | 0.013 | 0.829 |
| Switching capture error (RMSE) [19] | 0.013 | 0.039 | 0.327 |
| Switching capture error (RMSE) [10] | 0.007 | 0.049 | 0.148 |
| Physical limits | √ | × | × |
| Non-linear output | √ | × | × |

## V. CONCLUSION

In this study, we introduce the GEM model, a general and accurate memristive transistor model that effectively replicates the electrical characteristics of physical devices. The GEM model integrates a time-dependent differential equation, a moving window function, and a nonlinear output unit. These components collectively enable the GEM model to achieve an RMSE of less than 0.01 across various physical memristive transistor devices, representing a significant improvement of over 300% compared to existing SOTA models. In addition, this model faithfully implements the physical limits of state changes and nonlinear outputs observed in memristive transistors, with mean absolute errors of 0.041 $\mu$A and 0.017 $\mu$A, respectively. These results demonstrate the robust ability of the GEM model to accurately capture the electrical behavior of memristive transistors, significantly promoting the simulation and application of memristive transistors. The SPICE GEM model, along with the code and materials, are available at https://github.com/RTCartist/GEM-A-GEneral-Memristive-transistor-model.

## ACKNOWLEDGMENT

S.G. acknowledges funding from National Key Research and Development Program of China (grant No. 2023YFB3208003), National Natural Science Foundation of China (grant No. 62171014), and Beihang University (grants No. KG161250 and ZG16S2103).